\documentclass[conference]{IEEEtran}
\usepackage{setspace}
\usepackage{algorithm2e}
\usepackage{epsfig}
\usepackage{graphicx}
\usepackage{wrapfig}
\graphicspath {{Figures/}}\DeclareGraphicsExtensions{.eps,.ps}
\usepackage{comment}
\usepackage{subfigure}
\usepackage{multirow}
\usepackage{amssymb,amsmath}

\small\normalsize
\IEEEoverridecommandlockouts
\begin{document}
\pagestyle{plain}
\title{Network Coding-Based Link Failure Recovery over Large Arbitrary Networks}
\author{\IEEEauthorblockN{Serhat Nazim Avci and Ender Ayanoglu}\\
\IEEEauthorblockA{Center for Pervasive Communications and Computing\\
Department of Electrical Engineering and Computer Science\\
University of California, Irvine\\
Irvine, CA 92697-2625
}
\thanks{This work is partially supported by NSF under Grant No. 0917176.}
}
\maketitle
\begin{abstract}
%
\boldmath
Network coding-based link failure recovery techniques provide near-hitless recovery and offer high capacity
efficiency. Diversity coding is the first technique to incorporate coding in this field and is easy to implement
over small arbitrary networks. However, its capacity efficiency is restricted by its systematic coding and high design complexity even though it has lower complexity than the other coding-based recovery techniques. Alternative techniques mitigate some of these limitations, but they are difficult to implement over arbitrary networks. In this paper, we propose a novel non-systematic coding technique and a simple design algorithm to implement the diversity coding-based (or network coding-based) recovery over arbitrary networks. The design framework consists of two parts. An ILP formulation for each part is developed. The simulation results suggest that both the novel coding structure and the novel design algorithm lead to higher capacity efficiency for near-hitless recovery. The new design algorithm is able to achieve optimal results in large arbitrary networks.
\end{abstract} 
\section{Introduction}
The information carried by wide area networks is, in general, very important. Yet these networks regularly undergo failures. Detailed statistics about the network failures can be found in \cite{diversitysubMs}. This paper focuses on recovery from single link failures since they consist of 70\% of all network failures. To minimize the cost of such failures, various restoration and protection techniques are developed. The two main metrics in the design of these techniques are restoration speed and capacity efficiency. Capacity efficiency is measured by the total required capacity, in terms of fiber miles, and restoration speed is measured by the duration between the occurrence of failure and restoration of failed traffic. The goal is to minimize both of these metrics and every technique offers a different tradeoff.

In some recovery techniques, spare resources are shared among different traffic failure scenarios and different connection demands, whereas in others, spare resources are dedicated to connection demands. Dedicated protection techniques are able to offer near-hitless recovery since they do not require the signaling and rerouting of the failed traffic. 1+1 Automatic Protection Switching (APS) is a dedicated protection technique where two link-disjoint paths for each connection demand are employed to transmit the same data to the destination node. In the case of a link failure over the primary path, the destination node switches to the protection path and restores the traffic nearly instantaneously. However, 1+1 APS requires more than 100\% capacity which makes it capacity inefficient. The fact that 1+1 APS is currently employed in today's networks \cite{ibbarla} indicates the need for nearly instantaneous link failure recovery despite its low capacity efficiency.

The capacity efficiency of dedicated protection schemes such as 1+1 APS can be improved if the dedicated paths are shared. This can be achieved by employing coding, in particular, erasure coding \cite{aigm}, \cite{aigm2}. The technique introduced in \cite{aigm}, \cite{aigm2}, called diversity coding, has two advantages. First, unlike 1+1 APS, it is capacity efficient. Second, unlike rerouting-based restoration schemes, the recovery is nearly instantaneous. References \cite{aigm}, \cite{aigm2} predate network coding, usually considered to be introduced in \cite{acly}.

In \cite{diversity}, diversity coding is implemented over arbitrary networks using a heuristic algorithm. In \cite{diversitysubMs}, optimal algorithms for the diversity coding technique are developed. Diversity coding performs near-hitless recovery while offering competitive capacity efficiency. In \cite{CPP}, a solution of Shared Path Protection (SPP) \cite{rama} is converted to a coding-based solution named Coded Path Protection (CPP). Sharing of the spare resources is replaced with the employement of these resources to code different paths. This conversion increases the restoration speed and the transmission integrity, and decreases error signaling complexity. The bidirectional nature of CPP allows encoding and decoding inside the network for unicast demands.

In \cite{Kamal2} and \cite{Kamal3}, network coding-based protection schemes called 1+N protection are proposed in which coding operations are carried out over trees and trails, respectively. The idea is similar to that of diversity coding except the protection is bidirectional. In \cite{Hover}, the cost efficiencies of a network coding-based recovery technique and a simpler version of diversity coding technique are evaluated.

All of the above mentioned techniques implement systematic coding where coding operations are bound to specific protection topologies and primary paths are exempt from coding operations. In addition, they require strict link-disjointness between each primary path and the protection paths. Even though these assumptions make those techniques easier to implement, they have restricted capacity efficiencies.

In \cite{diversity_netcod}, the primary paths are incorporated into coding operations using a heuristic algorithm for static provisioning. The decodability of the coding structures is preserved by randomly adding the connection demands to the existing coding groups one by one. A coding group is a set of connection demands that are coded and protected together. Coding primary paths increases capacity efficiency over conventional diversity coding, as in \cite{diversity_netcod}.
Non-systematic coding is implemented in wireless mesh networks for single link failure recovery in \cite{kamal_many}. In \cite{rouy}, a general network-coding based approach is presented which employs non-systematic coding and does not explicitly require link-disjointness between primary paths and protection paths. However, this approach is restricted to specific topologies. In addition, it can protect at most two connection demands simultaneously. In \cite{rouyITA}, the proposed technique lifts the restriction over the number of protected connection demands for bidirected networks. 
In general, the coding-based recovery techniques in the literature, such as \cite{Kamal2}, \cite{rouy}, \cite{rouyITA}, offer promises in terms of capacity efficiency and restoration time. However, they cannot be optimally implemented on real networks due to their high design complexity limitations. The test networks and traffic matrices in those papers are much smaller than the real networks.

This paper offers two novel contributions to the field of diversity coding-based (or network coding-based) link failure recovery. First, we introduce an optimal, simple, and modular design algorithm that provisions the static traffic in relatively large networks. The underlying coding structure of this algorithm is arbitrary as long as the destination nodes of the connections are the same, which offers a solution for different techniques under the same framework. Second, we improve the coding structure of simple diversity coding by offering an optimal non-systematic coding structure using an Integer Linear Programming (ILP) formulation. In a non-systematic coding structure, both primary and protection paths are incorporated into the coding groups.
The performance of the new proposed coding technique is investigated compared to conventional (systematic) diversity coding using the novel design algorithm. The performance of the new design algorithm is also tested based on a set of simulations over a relatively large U.S. long-distance network.

\section{Design Algorithm}
The link failure recovery problem has two main components, namely an underlying recovery technique and a design algorithm. A recovery technique can achieve its potential performance only with a fast and optimal design algorithm that maps it over the networks of interest. Therefore, some recovery techniques can be theoretically advanced but they may perform poorly on test networks due to the high complexity of the accompanying design algorithm.

We developed a simple, optimal, and modular design algorithm for arbitrary single destination coding-based recovery techniques. The novelty is decomposing the design process into two parts, a pre-processing phase and the main problem solving phase. The design process is depicted in Fig.~\ref{fig:design}. The pre-processing phase has as its input the network graph and the destination node. In the pre-processing phase, all candidate coding groups are listed and their total cost to route and protect are calculated. These calculations depend on the underlying recovery technique. The size of a coding group is limited by the nodal degree of the destination node. If a coding group is not feasible, it has infinite total cost. The candidate coding group list is given as input to the main problem solving phase. In the main problem solving phase, those coding groups are optimally chosen and placed over the network such that all of the traffic demands are routed and protected through a coding group. The traffic matrix is decomposed into smaller vectors based on the destination nodes of the connections and input to the main problem. The main problem is inspired by the {\em p}-cycle algorithm in \cite{pcycle}. Both the coding groups formation and placement operations are carried out by ILP formulations which have dramatically fewer number of variables and constraints than those of \cite{diversitysubMs}, \cite{Kamal2}, and \cite{Hover}.

\ifCLASSOPTIONonecolumn
\begin{figure}[t!]
\centering
\includegraphics[bb = 79 119 655 404, width=120mm,clip=true]{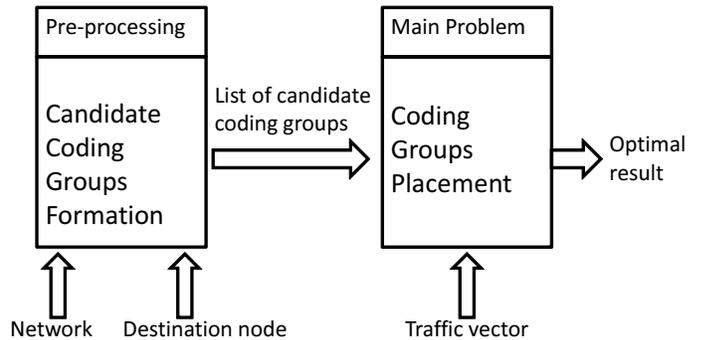}
\caption{Two phase design process leads to the optimal result.}
\label{fig:design}
\end{figure}
\else
\begin{figure}[!t]
\centering
\includegraphics[bb = 79 119 655 404,width=90mm,clip=true]{design_process.eps}
\caption{Two phase design process leads to the optimal result.}
\label{fig:design}
\end{figure}
\fi
For example, there is a network with three nodes, $S_1$, $S_2$, and $D$. Assume that there are 3 units and 2 units of traffic from $S_1$ and $S_2$ to $D$, respectively. In this scenario, there are three candidate coding groups with source nodes $[(S_1),(S_2),(S_1,S_2)]$. Assume that the total cost vector of these coding groups are $[5,10,12]$. The main problem inputs candidate coding groups with their cost vector and tries to minimize the total cost by satisfying the traffic demands. In the optimal solution, two units of coding group $(S_1,S_2)$ and one unit of coding group $(S_1)$ are placed over the network with minimum total cost 29.  

In \cite{PILP}, enumeration of {\em p}-cycles is argued to be slow. However, the nature of our problem is a better fit to enumeration of the candidate protection structures due to three reasons. First, the number of candidate coding groups is much smaller than the number of candidate {\em p}-cycles, when a single destination node is employed. Second, enumeration and placement of {\em p}-cycles are a spare capacity placement (SCP) operation and the routing of the primary paths must be handled separately. However, in our case, a coding group both routes and protects the connections, which is joint capacity placement (JCP). Third, the number of constraints in the {\em p}-cycle algorithm equals the number of edges, but the number of constraints in our algorithm is equal to the number of nodes.

The total number of candidate coding groups is an important criterion in terms of design complexity. For the typical coding-based recovery techniques, which employ a single destination node, the total number of candidate coding groups is proportional to
\begin{equation}
\dbinom{|V|-1}{ND-1} + \dbinom{|V|-1}{ND-2} + ... + \dbinom{|V|-1}{1},
\end{equation}
where $ND$ is the nodal degree and $|V|$ is the number of nodes in the network. $ND-1$ is the largest size of a coding group and $|V|-1$ defines the size of the list of source nodes a coding group choose from. $ND$ is the most important parameter defining the complexity of the new algorithm. On the other hand, the size of the traffic matrix is negligible in terms of complexity since traffic demands only take place in the right hand side of the constraints of the ILP formulation of the main problem. They do not affect the number of variables or constraints in the ILP formulation. We want to make the important point that the proposed algorithm is also robust to changes in the traffic matrix. If the traffic matrix changes over time, there is no need to carry out the pre-processing phase again. The right hand side of the constraints in the main problem can be changed to optimize the network in response to changing traffic. The main problem is very fast since it has only $|V|-1$ constraints.  
\section{Non-systematic Coding}

\label{sec:nonsystem}
In this section, we introduce an ILP-based optimal non-systematic diversity coding structure for single link failure recovery. 

We assume that the connection demands in the same coding group have a common destination node. Their source nodes can be the same or different. There are $N$ connection demands in a coding group. Each connection demand has two link-disjoint paths carrying the same signal, which is distinct from other connection demands. Some of these paths are combined and coded together and some of them are not combined with any other path. For simplicity, we assume all of the operations are over $GF(2)$, although this assumption can be relaxed, e.g., \cite{aigm}, \cite{aigm2}. The paths in a coding group are assigned to subgroups. The total number of subgroups varies between $N+1$ and $2N$. The number of paths in a subgroup take values from zero to $N$. The paths in the same subgroup are assumed to be coded together. In the received vector of the destination node, each connection demand is represented as a variable and each subgroup is represented as an equation. Clearly, if there are smaller than or equal to $N$ subgroups, some data cannot be recovered in some failure scenarios because that leaves $N-1$ equations for $N$ unknowns. In the opposite extreme, there will be a maximum $2N$ subgroups if each path is transmitted separately, which is the case in 1+1 APS.

In conventional diversity coding against single link failures, there exists $N$ primary paths, one for each connection demand, and a single protection path carrying the modulo-2 sum of the data over primary paths. Each connection demand is delivered to the destination node over two link-disjoint paths. It has a total of $N+1$ subgroups, $N$ of them are the primary paths and one of them is the combination of protection paths. The protection path topology can be a tree if it is formed by combining paths originating from different source nodes. The coding operations are restricted over the protection path (tree). The common destination node carries out the decoding operation over the received vector. The example in Fig.~\ref{fig:capExample}, taken from \cite{diversity_netcod}, shows how non-systematic coding can reduce the total capacity for the protection of a coding group. The paths in a non-systematic code are equivalent to each other and therefore cannot be categorized as primary and protection paths. There are four connection demands destined to node $D$. Two of them are originated from S1, represented by symbols $a$ and $b$. The other two are originated from $S2$ and $S3$, represented by symbols $c$ and $d$, respectively. All four connection demands form a coding group. In Fig.~\ref{fig:subfig1}, a typical diversity coding solution is depicted. The common protection path is shown with dashed lines. In Fig.~\ref{fig:subfig2}, a non-systematic coding solution is depicted. It enables what was once protection path of $c$ to be coded with what was once primary path of $b$ over nodes $4-5$. That coding operation eliminates the need for the link between $4-1$ carrying $c$. Therefore, non-systematic coding can improve the capacity efficiency. In the worst case scenario, it performs the same as systematic conventional diversity coding.

\ifCLASSOPTIONonecolumn
\begin{figure}[m!]
\centering
\subfigure[]{
\includegraphics[bb = 90 177 565 460,width = 60mm]{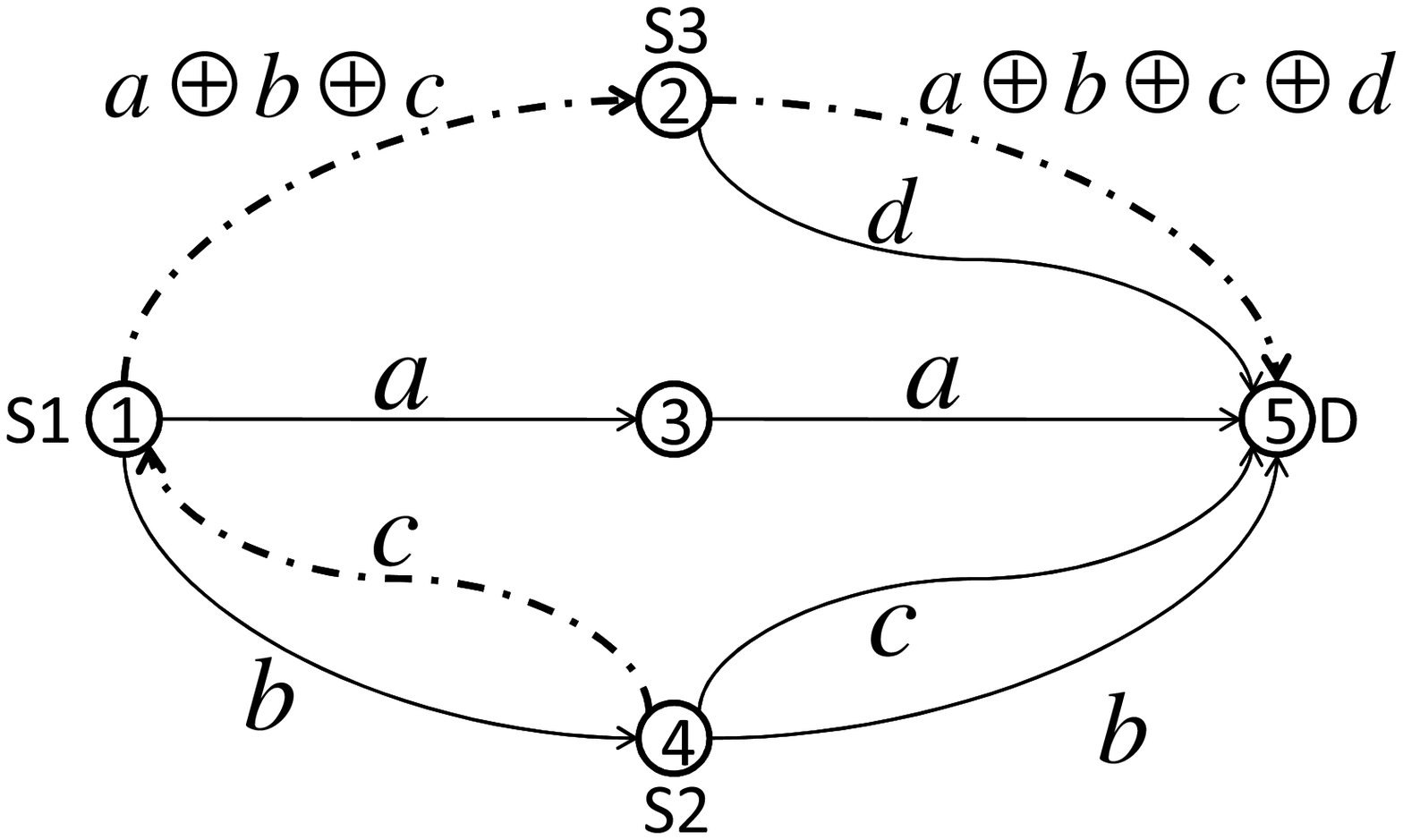}
\label{fig:subfig1}
}
\subfigure[]{
\includegraphics[bb = 90 177 565 460,width = 60mm,clip=true]{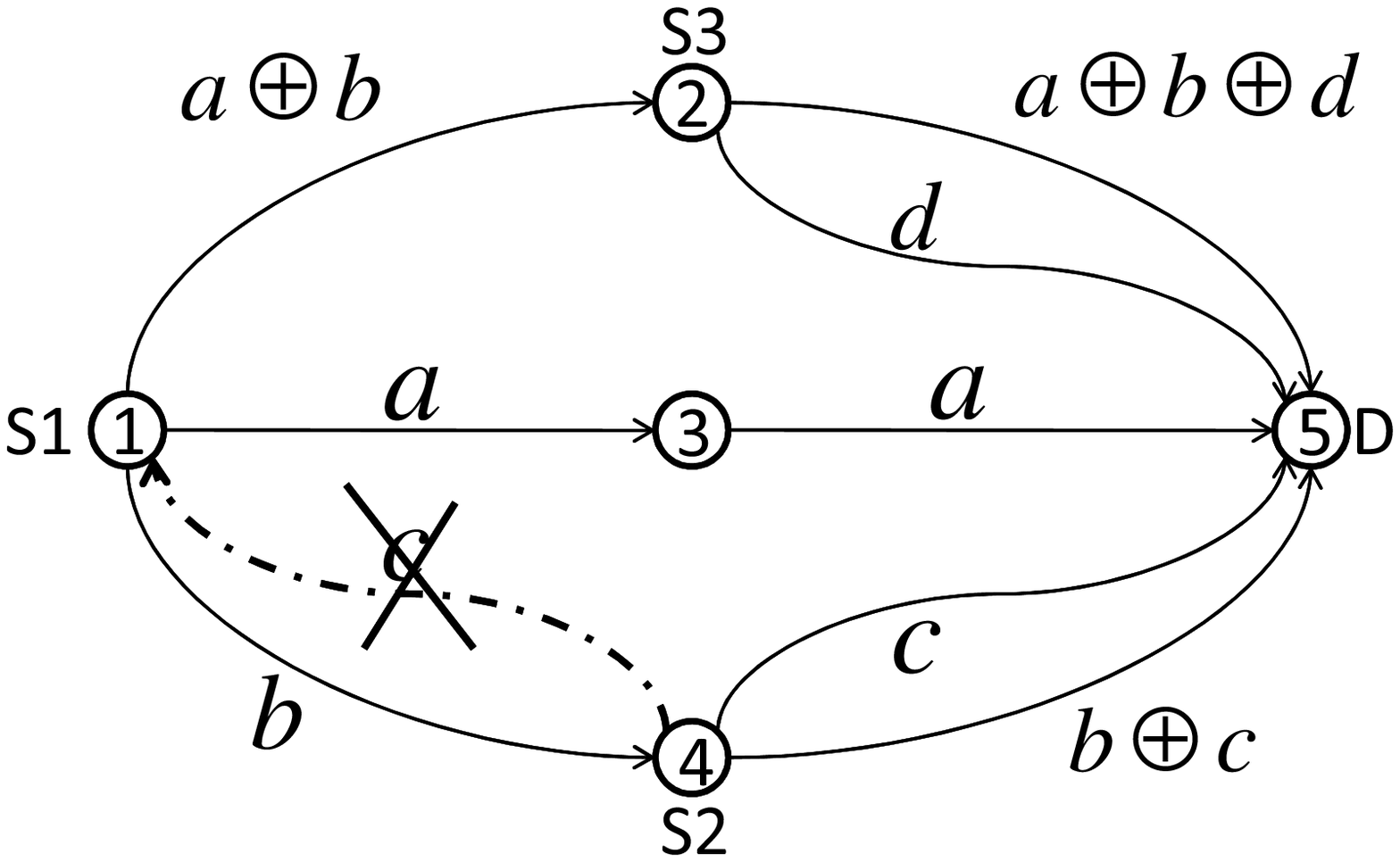}
\label{fig:subfig2}
}
\caption[Optional caption for list of figures]{Capacity efficiency improvement with non-systematic coding \subref{fig:subfig1} Diversity coding solution, \subref{fig:subfig2} Non-systematic coding solution}
\label{fig:capExample}
\end{figure}
\else
\begin{figure}[!t]
\centering
\subfigure[]{
\includegraphics[bb = 90 177 565 460,width = 60mm]{Capacity_Example1.eps}
\label{fig:subfig1}
}
\subfigure[]{
\includegraphics[bb = 90 177 565 460,width = 60mm,clip=true]{Capacity_Example2.eps}
\label{fig:subfig2}
}
\caption[Optional caption for list of figures]{Capacity efficiency improvement with non-systematic coding, \subref{fig:subfig1}Diversity coding solution, \subref{fig:subfig2} Non-systematic coding solution}
\vspace{-2mm}
\label{fig:capExample}
\end{figure}
\fi

A non-systematic code can be built by assigning paths to the subgroups arbitrarily. However, the critical point in the construction of a non-systematic code is the decodability of all $N$ transmitted signals.  The $N$ data signal can be decoded under any single link failure scenario as long as any $N$ equations of the received vector are linearly independent. It is clear that any subset of linear equations with size $N$ of the received vector are independent in systematic diversity coding. The received vector of systematic diversity coding for four connection demands is
\begin{equation}
\left[\begin{tabular}{c}
$a$\\
$b$\\
$c$\\
$d$\\
$a+b+c+d$\\
\end{tabular}\right],
\end{equation}
where $a$, $b$, $c$, and $d$ are transmitted signals by each connection demand. $N+1$ subgroups are sufficient for systematic diversity coding. In non-systematic coding, the paths in each subgroup must be specified. In \cite{diversity_netcod}, connection demands are randomly chosen and paths are assigned one by one to subgroups of the existing coding groups. However, a general rule is needed to optimally build non-systematic codes. In \cite{kamal_many}, it is reported by \textit{Lemma 1} that the destination node can recover $N$ data signals from a non-systematic code as long as any subset of the data signals with size $k$ are transmitted over at least $k+1$ paths. In our technique, \textit{Lemma 1} changes to

\textbf{\textit{Lemma 1.}} \textit{The non-systematic code will be valid as long as any subset of data signals with size $k$ are members of at least $k+1$ subgroups in a coding group.}

The proof follows from \cite{kamal_many}, assuming $U_s$ as the set of connection demand signals and $L_s$ as the set of subgroups in a coding group.

This paper aims to build valid non-systematic codes with the objective of minimizing total capacity. Therefore, we develop an optimization algorithm to find the code that requires lowest total capacity while eliminating the codes that violate \textit{Lemma 1}.
The following example shows how an invalid non-systematic code can be detected. Assume we have four connection demands, carrying signals $a$, $b$, $c$, and $d$ in a coding group and each connection demand has two link-disjoint paths. Assume the first three subgroups of this coding group are given as
\begin{equation}
\left[\begin{tabular}{c}
$a+b$\\
$b+c$\\
$c+d$\\
\end{tabular}\right],
\end{equation}
which indicate that one path of $a$ and $b$, $b$ and $c$, and $c$ and $d$ are coded together. That leads to a coding relationship map shown in Fig.~\ref{fig:cod1}. In this map, there are two symbols for each connection demand, referring to their two link-disjoint paths. In Fig.~\ref{fig:cod2}, a bidirectional arrow between two paths means they are in the same subgroup and therefore coded together. If a path of $a$ is coded together with a path of $b$ and a path of $b$ is coded together with a path of $c$, then connection demand $a$ is indirectly related to connection demand $c$, which is shown with a dashed arrow in Fig.~\ref{fig:cod2}. In addition, pairs $a-d$ and $b-d$ are indirectly related as well. If the fourth subgroup consists of $a+d$ then four connection demands are bounded within four subgroups, which is a violation of \textit{Lemma 1}. In Fig.~\ref{fig:cod3}, the relationship map is updated to include a bidirectional arrow between a path of $a$ and a path of $d$. As a result, connection demands $a$ and $d$ are coded together and indirectly related at the same time, which causes a circle shown in Fig.~\ref{fig:cod4}. We call this a coding circle, which is an indication of the violation of \textit{Lemma 1}. Therefore, in the ILP formulation, we seek to prevent coding circles by ensuring two different connection demands can either be coded together or are indirectly related. The resulting non-systematic code will be valid as long as coding circles are prevented.

\ifCLASSOPTIONonecolumn
\begin{figure}[m!]
\centering
\subfigure[]{
\includegraphics[width = 58mm]{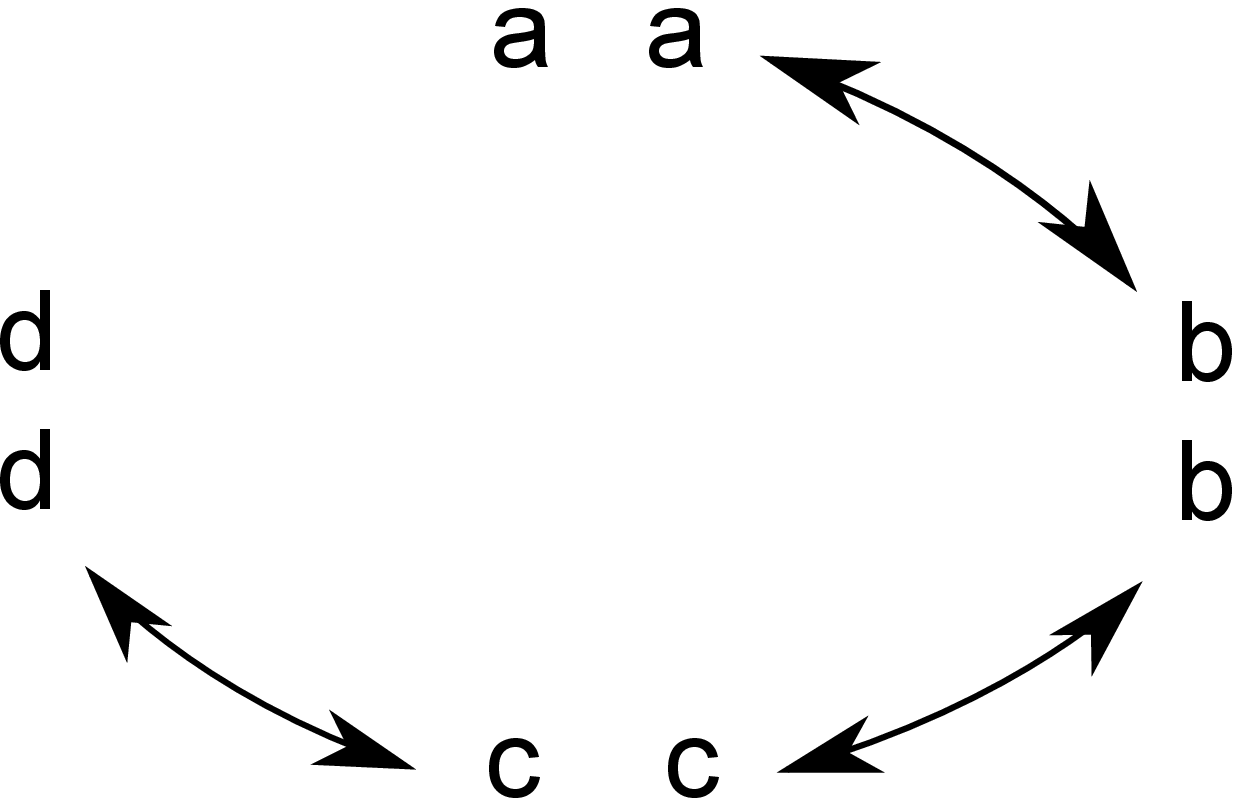}
\label{fig:cod1}
}
\subfigure[]{
\includegraphics[width = 60mm]{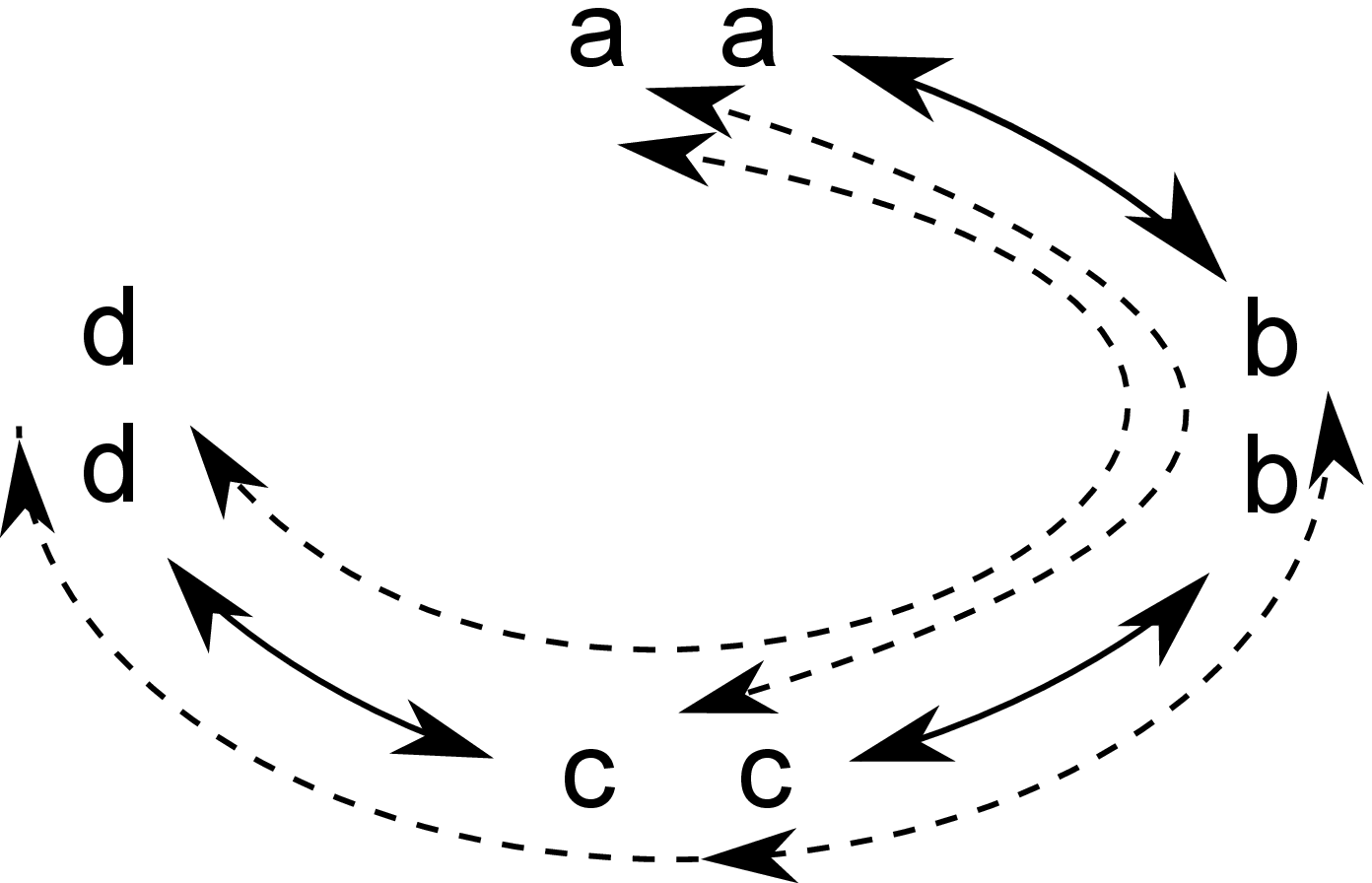}
\label{fig:cod2}
}
\subfigure[]{
\includegraphics[width = 60mm]{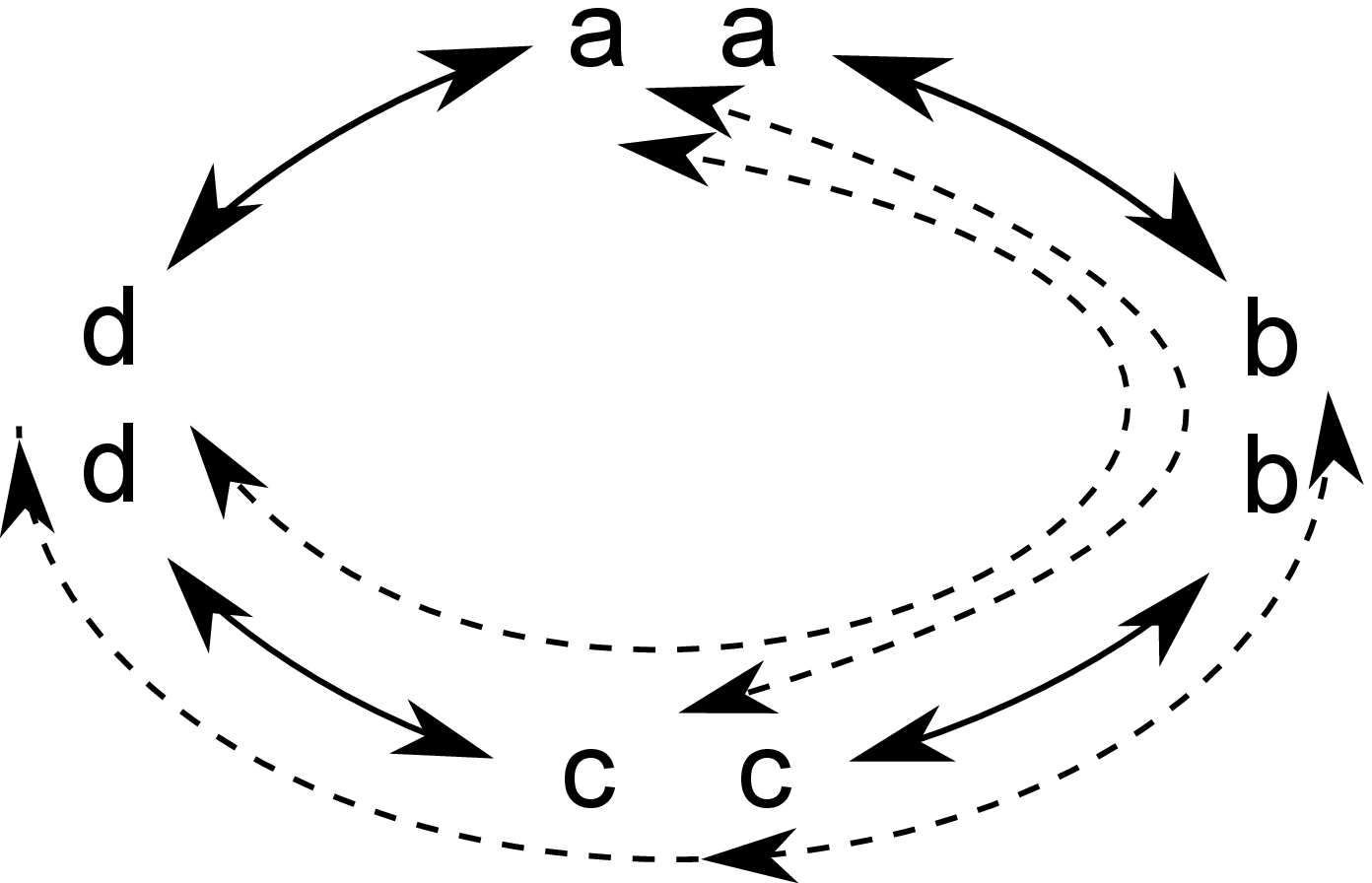}
\label{fig:cod3}
}
\subfigure[]{
\includegraphics[width = 60mm]{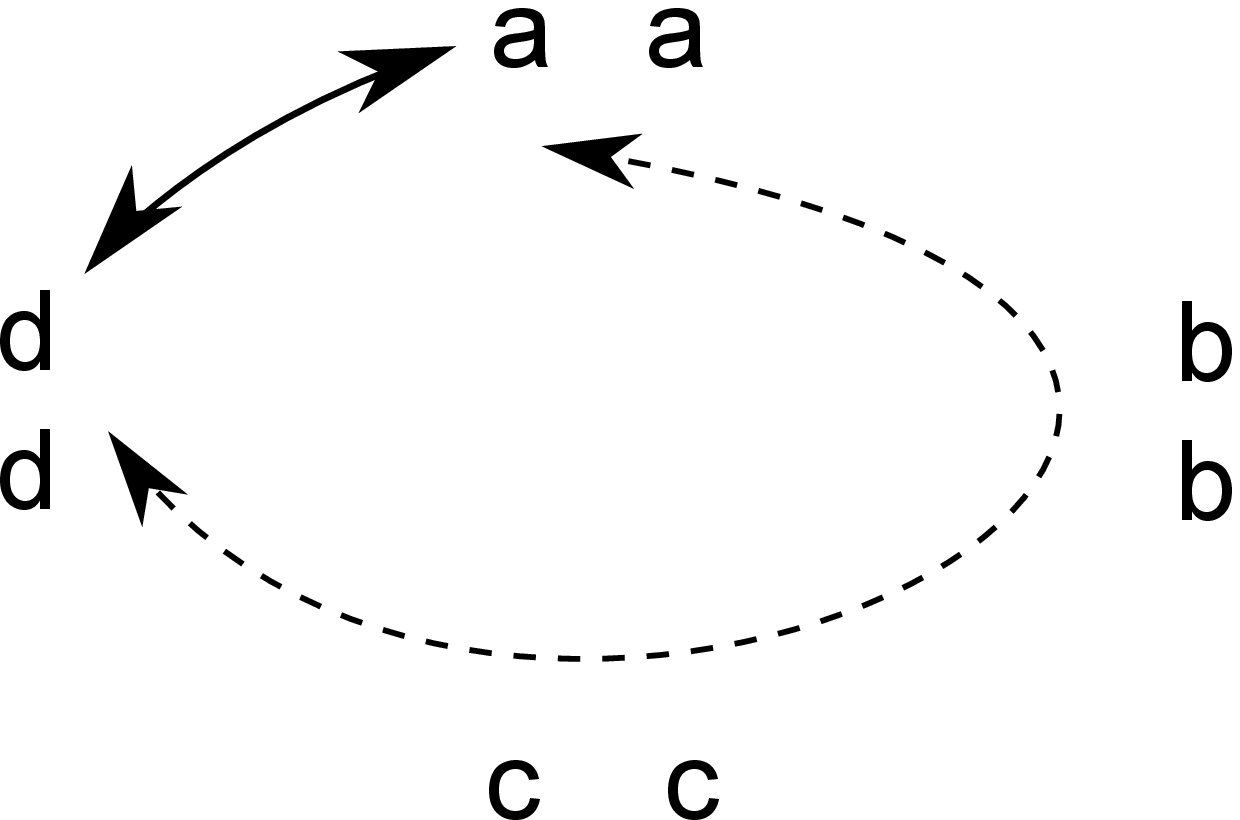}
\label{fig:cod4}
}
\label{fig:codeExample}
\caption[Optional caption for list of figures]{Formation of a coding circle. A coding circle violates \textit{Lemma 1}.}
\end{figure}
\else
\begin{figure}[!t]
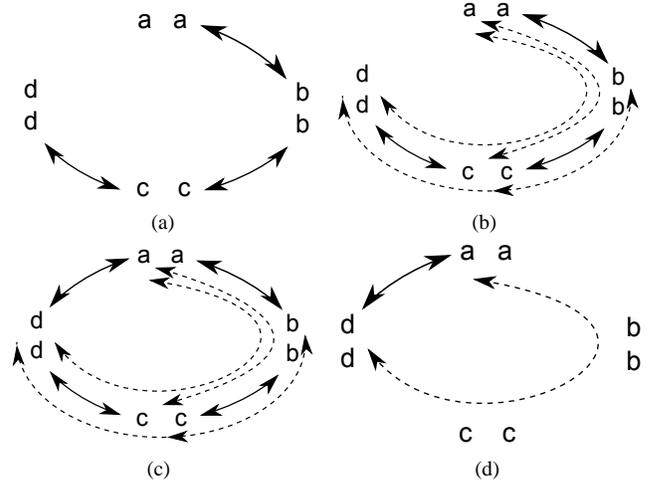

\centering
\subfigure[]{
\includegraphics[width = 38mm]{Coding_circle1.eps}
\label{fig:cod1}
}
\subfigure[]{
\includegraphics[width = 40mm]{Coding_circle2.eps}
\label{fig:cod2}
}
\subfigure[]{
\includegraphics[width = 40mm]{Coding_circle3.eps}
\label{fig:cod3}
}
\subfigure[]{
\includegraphics[width = 40mm]{Coding_circle4.eps}
\label{fig:cod4}
}
\label{fig:codeExample}
\caption[Optional caption for list of figures]{Formation of a coding circle. A coding circle violates \textit{Lemma 1}.}
\vspace{-2mm}
\end{figure}
\fi

\section{ILP Formulations}
\subsection{Candidate Coding Groups Formation}
\label{sec:Nonsys}
An ILP formulation is developed to implement the proposed technique with an objective to minimize the total capacity (cost) of a coding group in arbitrary networks. The ILP formulation finds the optimum non-systematic diversity coding structure by simply going through all possible subgroup assignments for each path and eliminating the ones which violate \textit{Lemma 1}. The input parameters of the ILP are

\begin{itemize}
\item $G(V,E)$   : Network graph,
\item $S$        : The set of spans in the network, there are two opposite directional links in each span,
\item $N$        : Enumerated list of all connections,
\item $P$        : Enumerated list of all paths, $\left|{P}\right|=2 \times \left|{N}\right|$,
\item $a_e$      : Cost associated with link $e$,
\item $\Gamma_i(v)$ : The set of incoming links of each node $v$,
\item $\Gamma_o(v)$ : The set of outgoing links of node $v$,
\item $s_i$      : The source node of path $i$,
\item $d$        : The destination node.
\end{itemize}
The variables related to finding two paths for each connection are
\begin{itemize}
\item $x_e(i)$   : Equals 1 iff the path $i$ passes through link $e$, 0 otherwise.
\end{itemize}
The following inequality finds two paths for each connection demand
\begin{equation}
\sum_{e\in\Gamma_i(v)}x_e(i)-\sum_{e\in\Gamma_o(v)}x_e(i)=\left\{\begin{tabular}{ll}
-1&{\rm if\ }$v=s_i,$\\
1&{\rm if\ }$v=d,$\\
0&{\rm otherwise.}\end{tabular}
\right.
\end{equation}
Note that we require $\bmod{(i,2)}=0 \Rightarrow s_i=s_{i-1}$ for $1\leq i \leq 2N$.
The variables related to finding a valid non-systematic code are
\begin{itemize}
\item $n(i,s)$   : Equals 1 iff path $i$ is in subgroup $s$, 0 otherwise,
\item $m(i,j)$   : Equals 1 iff path $i$ and path $j$ are in the same subgroup so are coded together, 0 otherwise,
\item $r(i,f)$   : Equals 1 iff path $i$ and connection demand $f$ are indirectly related, 0 otherwise.
\end{itemize}
Each path must be assigned to a single subgroup which is ensured by
\begin{equation}
\sum_{s=1}^{2N} n(i,s)=1 \hspace{5mm}\forall i,
\end{equation}
\begin{equation}
n(i,s) + n(i-1,s) \leq 1 \hspace{5mm} \forall i,s : \bmod(i,2)=0, \label{ineq:3}
\end{equation}
\begin{equation}
m(i,j) \geq n(i,s) + n(j,s)-1 \hspace{5mm} \forall i\neq j,s.  \label{ineq:4}
\end{equation}
Inequality (\ref{ineq:3}) ensures that complementary paths cannot be in the same subgroup. If two paths are in the same subgroup, then they are assumed to be coded together, which is satisfied by inequality (\ref{ineq:4}).
\begin{eqnarray}\nonumber
r(i,f) \geq m(i,j)+m(j^*,2f)+m(j^*,2f-1)\\
-m(i,2f) -m(i,2f-1)-1 \hspace{5mm} \forall i,j,f,:i\neq j \label{ineq:5}
\end{eqnarray}
where $j^*=j-1$ if $\bmod(j,2)=0$ and $j^*=j+1$ otherwise.
\begin{eqnarray}\nonumber
r(i,f) \geq r(i,g)+m(2g,2f)+m(2g-1,2f-1)\\
\nonumber +m(2g-1,2f) +m(2g-1,2f-1)-1 \hspace{2mm} \forall i,f\neq g: \\
i\neq 2f, i\neq 2f-1,i\neq 2g, i\neq 2g-1, \label{ineq:6}
\end{eqnarray}
\begin{eqnarray}\nonumber
r(2f,g) + r(2f-1,g)+m(2f,2g)+m(2f-1,2g) \\
+m(2f,2g-1) +m(2f-1,2g-1)\leq 1 \hspace{2mm} \forall g,f : g\neq f. \label{ineq:7}
\end{eqnarray}
In inequality (\ref{ineq:5}), path $i$ becomes indirectly related to demand $f$ if path $i$ is coded with path $j$ and if there exists a path j that is coded with both path $i$ and one of the paths carrying demand $f$. Moreover, path $i$ must not be coded with either paths of demand $f$. Inequality (\ref{ineq:6}) ensures that path $i$ becomes indirectly related to demand $f$ if path $i$ is indirectly related to demand $g$ and one of the paths carrying demand $g$ is coded with one the paths carrying demand $f$. 
Inequality (\ref{ineq:7}) ensures that only one of the paths carrying demand $f$ can be either coded with one of the paths carrying demand $g$ or be indirectly related to demand $g$. This inequality ensures the validity of the non-systematic code by preventing coding circles.
The final variable of the ILP is
\begin{itemize}
\item $t_e(s)$   : Equals 1 iff one of the paths in subgroup $s$ traverses over link $e$, 0 otherwise.
\end{itemize}
\begin{equation}
t_e(s)\geq x_e(i)+n(i,s)-1 \hspace{5mm} \forall e,i,s.
\label{ineq:11}
\end{equation}
\begin{eqnarray}\nonumber
t_e(s_1) + t_e(s_2) + t_f(s_1) + t_f(s_2) \leq 1\\
\forall e,f \in g, \forall g\in S, \forall s_1,s_2 \label{ineq:12}
\end{eqnarray}
Inequality (\ref{ineq:11}) finds the topology of each subgroup. The topology of a subgroup is the union of the protection paths of the connections in that subgroup. Inequality (\ref{ineq:12}) ensures that the topologies of two subgroups are link-disjoint.

The objective function is
\begin{equation}
\min \sum_{e\in E}\sum_{s=1}^{2N} t_e(s)\times a_e.
\end{equation}

\subsection{Coding Groups Placement}
We assume that there is a single destination node and the rest are source nodes. There is a single connection demand from each source node to the destination with varying traffic rates. Those connection demands can be split into unit demands and protected via different coding groups.
In addition to the parameters in the previous section, the extra parameters are
\begin{itemize}
\item $CG$      : The candidate coding groups list,
\item $CG_{i,f}$  : Equals 1 iff the candidate coding group $i$ includes connection demand $f$, 0 otherwise
\item $c_i$      : Total cost (capacity) of coding group $i$,
\item $t_f$      : Traffic rate of the connection demand $f$.   
\end{itemize} 
We have only one set integer variables 
\begin{itemize}
\item $n_i$ : The number of units of coding group $i$ that are placed. 
\end{itemize}
The objective function is 
\begin{equation}
\min \sum_{i=1}^{|CG|} c_i \times n_i
\end{equation}
subject to 
\begin{equation}
\sum_{i=1}^{|CG|} CG_{i,f}\times n_i \geq t_f \hspace{8mm} 1\leq f\leq |V|-1, 
\end{equation}
which ensures that the placed coding groups are sufficient to cover the traffic demands. Even though the 
coding group placement problem may have a high number of variables in large networks, the fact that it only has $|V|-1$ constraints makes it achieve the optimal results in sub-ms.

\section{Simulation Results}
\ifCLASSOPTIONonecolumn
\begin{table*}[h!]
\centering
\caption{SCaP Results for Each Destination Node (\%)}
\begin{tabular}{|l|c|c|c|c|c|c|}
\hline
\multirow{3}{*}{Destination Node}
&\multicolumn{2}{|c|}{Diversity Coding Tree}
&\multicolumn{4}{|c|}{Coding Groups Placement Algorithm}\\ \cline{4-7}
&\multicolumn{2}{|c|}{}&
\multicolumn{2}{|c|}{Systematic Diversity Coding}&
\multicolumn{2}{|c|}{Non-systematic Diversity Coding}\\\cline{2-7}
&SCaP &Optimality gap&SCaP&Optimality gap&SCaP&Optimality gap \\ \hline
Node 1&84.1&24.9&79.0&0.0&74.4&0.0 \\ \hline
Node 2&74.8&23.3&69.5&0.0&68.0&0.0 \\ \hline
Node 3&65.6&14.8&64.3&0.0&62.5&0.0 \\ \hline
Node 4&88.7&22.8&87.6&0.0&82.9&0.0 \\ \hline
Node 5&80.4&25.2&69.2&0.0&63.5&0.0 \\ \hline
Node 6&91.5&21.3&84.7&0.0&74.1&0.0 \\ \hline
Node 7&95.0&20.8&89.6&0.0&85.8&0.0 \\ \hline
Node 8&106.1&19.8&99.9&0.0&90.7&0.0 \\ \hline
Node 9&87.8&15.0&85.7&0.0&82.7&0.0 \\ \hline
Node 10&116.7&24.9&106.6&0.0&96.3&0.0 \\ \hline
Node 11&92.4&18.2&85.7&0.0&77.1&0.0 \\ \hline
Average&87.9&21.5&82.1&0.0&76.9&0.0 \\ \hline
\end{tabular}
\label{table-CapEff}
\end{table*}
\else
\begin{table*}[t]
\centering
\caption{SCaP Results for Each Destination Node}
\begin{tabular}{|l|c|c|c|c|c|c|}
\hline
\multirow{3}{*}{Destination Node}
&\multicolumn{2}{|c|}{Diversity Coding Tree}
&\multicolumn{4}{|c|}{Coding Groups Placement Algorithm}\\ \cline{4-7}
&\multicolumn{2}{|c|}{}&
\multicolumn{2}{|c|}{Systematic Diversity Coding}&
\multicolumn{2}{|c|}{Non-systematic Diversity Coding}\\\cline{2-7}
&SCaP(\%) &Optimality gap&SCaP(\%)&Optimality gap&SCaP(\%)&Optimality gap \\ \hline
Node 1&84.1&24.9&79.0&0.0&74.4&0.0 \\ \hline
Node 2&74.8&23.3&69.5&0.0&68.0&0.0 \\ \hline
Node 3&65.6&14.8&64.3&0.0&62.5&0.0 \\ \hline
Node 4&88.7&22.8&87.6&0.0&82.9&0.0 \\ \hline
Node 5&80.4&25.2&69.2&0.0&63.5&0.0 \\ \hline
Node 6&91.5&21.3&84.7&0.0&74.1&0.0 \\ \hline
Node 7&95.0&20.8&89.6&0.0&85.8&0.0 \\ \hline
Node 8&106.1&19.8&99.9&0.0&90.7&0.0 \\ \hline
Node 9&87.8&15.0&85.7&0.0&82.7&0.0 \\ \hline
Node 10&116.7&24.9&106.6&0.0&96.3&0.0 \\ \hline
Node 11&92.4&18.2&85.7&0.0&77.1&0.0 \\ \hline
Average&87.9&21.5&82.1&0.0&76.9&0.0 \\ \hline
\end{tabular}
\label{table-CapEff}
\end{table*}
\fi
In this section, we present two different simulations to investigate the performance of the proposed coding technique and the proposed design algorithm differentially. The first test network is COST 239 network, which is depicted in Fig.~\ref{COST239 network}. There are 3 units of uniform traffic between each node pair. The performance metric is the spare capacity percentage (SCaP) as defined in \cite{diversity}. The goal is to measure the decrease in SCaP due to the introduction of non-systematic diversity coding. We also investigate how the new simplified design algorithm enables us to achieve better (optimal) results for systematic diversity coding than a competitive technique. The competitive technique is chosen as diversity coding tree algorithm in \cite{diversitysubMs} because it requires fewer number of variables and constraints than \cite{Kamal2} and \cite{Hover}. CPLEX 12.2 is used for the simulations.

\ifCLASSOPTIONonecolumn
\begin{figure}[t!]
\centering
\includegraphics[bb= 160 75 600 447,width=60mm,clip=true]{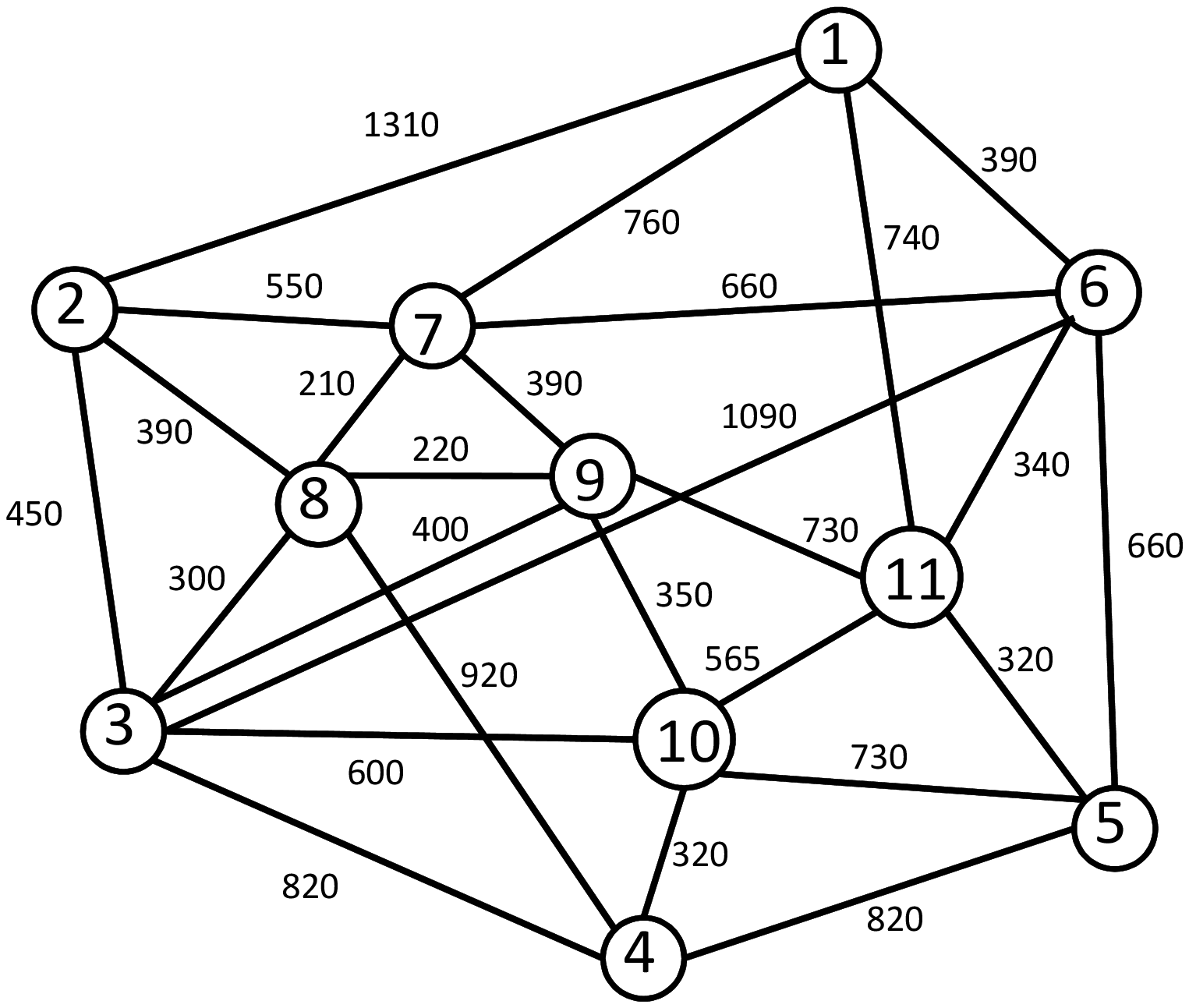}
\caption{European COST 239 network.}
\label{COST239 network}
\end{figure}
\else
\begin{figure}[t!]
\centering
\includegraphics[bb= 160 75 600 447,width=60mm,clip=true]{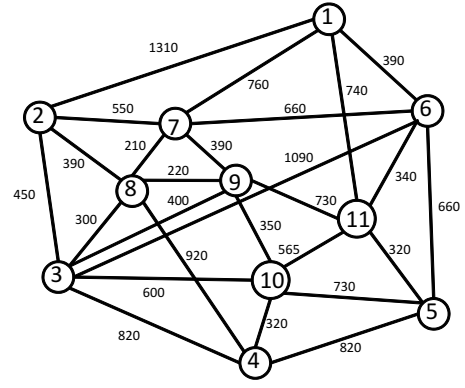}
\caption{European COST 239 network.}
\label{COST239 network}
\end{figure}
\fi

The SCaP and optimality gap values of three different schemes are shown in Table ~\ref{table-CapEff} in terms of percentiles. Systematic diversity coding via the new algorithm is derived by putting the secondary paths to the same subgroup in Section \ref{sec:Nonsys}. Maximum number of candidate coding groups is equal to 3002 when the destination node is equal to 3, which has the highest nodal degree.

The simulation results highlight three important points. First, the proposed algorithm can achieve the optimal result in all cases, whereas diversity coding tree algorithm tackles with memory limitations and cannot achieve optimal results. Second, even though the same systematic diversity coding is employed in the first and second algorithms, the proposed algorithm achieves better results since it can find the optimal result before the simulation terminates. It is noteworthy that the more the optimality gap of the first algorithm, the more difference between the SCaP values of two different algorithms. This clarifies the importance of the design algorithm for a recovery technique to achieve the promised results. The third point is the improvement in SCaP results due to the improvement in the coding structure. For all cases, non-systematic diversity coding performs better than its systematic counterpart. The difference in terms of SCaP exceeds 10\% in some scenarios.

\ifCLASSOPTIONonecolumn
\begin{table*}[h!]
\centering
\caption{SCaP Results for Each Destination Node (\%)}
\begin{tabular}{|l|c|c|c|}
\hline
Protection Technique&SCaP&Design complexity&Optimization Type \\ \hline
{\em P}-cycle algorithm \cite{GroverBook}&107.0\%&$\geq$ 1000000 ({\em p}-cycles)&SCP \\ \hline
Coding groups placement for systematic diversity coding&95.4\%&31464 (coding groups)&JCP \\ \hline
\end{tabular}
\label{table-Large}
\end{table*}
\else
\begin{table*}[t]
\centering
\caption{Comparative performance of the new algorithm in U.S. long-distance network}
\begin{tabular}{|l|c|c|c|}
\hline
Protection Technique&SCaP&Design complexity&Optimization Type \\ \hline
{\em P}-cycle algorithm \cite{GroverBook}&107.0\%&$\geq$ 1000000 ({\em p}-cycles)&SCP \\ \hline
Coding groups placement for systematic diversity coding&95.4\%&31464 (coding groups)&JCP \\ \hline
\end{tabular}
\label{table-Large}
\end{table*}
\fi
The second test network is the U.S. long-distance network, taken from \cite{XM99}, which is shown in Fig.~\ref{fig:USlong}. The traffic matrix is created using a gravity-based model \cite{zrdg}. In total, there are 23,204 static unit connection demands. This setup is chosen in order to observe the performance of the new design algorithm in a large realistic network with a dense traffic scenario. The other coding-based recovery design algorithms are too complex to implement in this setup. The SCaP results and the complexity of the algorithm are compared with a {\em p}-cycle algorithm in \cite[p. 699]{GroverBook}, which is considered to be within 5\% of the optimal solution. Both of the algorithms have a pre-processing phase where they enumerate all the candidate {\em p}-cycles or candidate coding groups. The results are presented in Table~\ref{table-Large}.
\ifCLASSOPTIONonecolumn
\begin{figure}[m!]
\centering
\includegraphics[width=80mm]{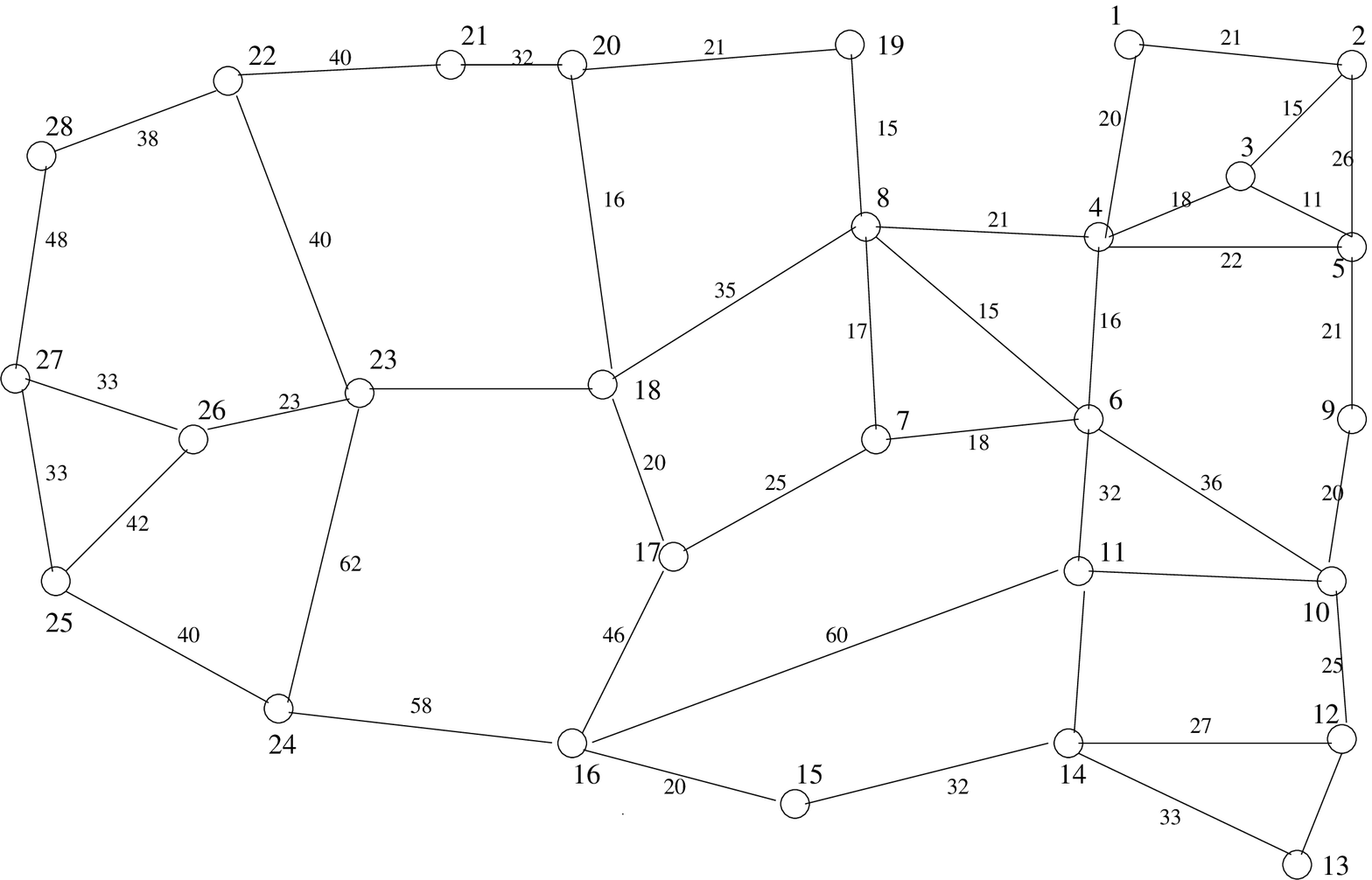}
\caption{U.S. long distance network.}
\label{fig:USlong}
\end{figure}
\else
\begin{figure}[!t]
\centering
\includegraphics[width=80mm]{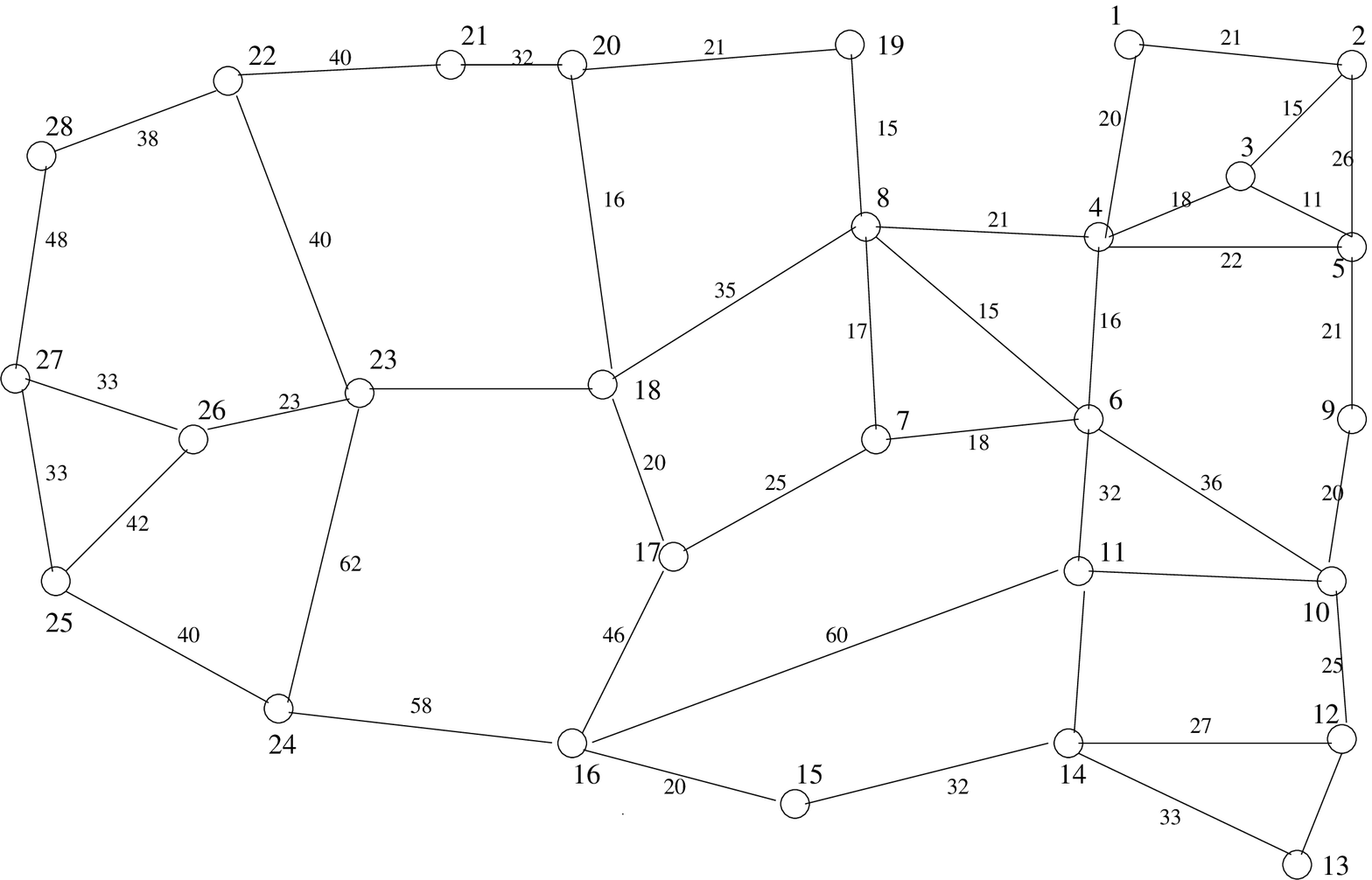}
\caption{U.S. long distance network.}
\label{fig:USlong}
\end{figure}
\fi

As seen from the results, the proposed design algorithm can achieve optimal results with conventional diversity coding even in a large realistic network with a dense traffic scenario. The SCaP result of the new technique is better than that of the {\em p}-cycle algorithm. It should be noted that, {\em p}-cycle algorithm carries out SCP, whereas the proposed algorithm carries out JCP. One important difference is the design complexity between two techniques, which adopts a similar idea. In the U.S. long-distance network, the number of candidate cycles is much higher than the number of candidate coding groups because the latter is constrained by the nodal degree of the network. As a future work, we plan to employ column generation technique, as done for {\em p}-cycles in \cite{Colu}, to make our algorithm even more scalable in larger networks.  
\section{Conclusion}

In this paper, we introduced an ILP-based non-systematic coding approach and a simple design algorithm to achieve near instantaneous recovery with higher capacity efficiency. Non-systematic coding allows any path in the coding group to be coded with other paths without compromising the decodability at the destination node. The code is developed with the objective of minimum capacity. These two advanced techniques combined achieve results with higher capacity efficiency. The advantages of both techniques are shown with examples and simulation results.

The new design framework consist of two parts, a pre-processing phase where the candidate coding groups are formed and the main problem solving phase where the optimal coding groups are placed over the network. We have developed an ILP formulation for each of these steps. In the pre-processing phase, coding groups are formed under the optimal non-systematic diversity coding principles. The main problem consists of only $|V|-1$ constraints. It finds and places the optimal coding group combinations to match the traffic demands, which takes sub-ms to run. The new algorithm can be implemented over networks with arbitrary topology and it can achieve optimal results in large networks for arbitrary traffic scenarios.

We ran two sets of simulations. Non-systematic diversity coding has a better capacity efficiency than conventional systematic diversity coding. In addition, we observe the significance of the simplicity of the design algorithm to achieve better results. In the later simulations, coding group placement algorithm is compared to a similar algorithm employing {\em p}-cycle protection over realistic U.S. long-distance network. The proposed algorithm achieves the optimal result.

%
%

\vspace{-1.5mm}
\bibliographystyle{IEEEtran}
\bibliography{IEEEabrv,bibliography/ISIT}
\end{document}